\documentclass[aps,prm,twocolumn,amsmath,amssymb,floatfix,superscriptaddress,citeautoscript]{revtex4-2}

\usepackage{graphicx}
\usepackage{natbib}
\usepackage{color}
\usepackage{hyperref}

\begin{document}

	\title{Topological Hall effect in CeAlGe}
	
	\author{M.~M.\ Piva}
	\email{Mario.Piva@cpfs.mpg.de}
	\affiliation{Max Planck Institute for Chemical Physics of Solids, N\"{o}thnitzer Stra{\ss}e 40, D-01187 Dresden, Germany}
	
	\author{J.~C.\ Souza}
    \altaffiliation[Present address: ]
    {Department of Condensed Matter Physics, Weizmann Institute of Science, Rehovot, Israel.}
    \affiliation{Instituto de F\'{\i}sica ``Gleb Wataghin'', UNICAMP, 13083-859, Campinas, SP, Brazil}
    \affiliation{Max Planck Institute for Chemical Physics of Solids, N\"{o}thnitzer Stra{\ss}e 40, D-01187 Dresden, Germany}
	
	\author{G.~A.\ Lombardi}
	\affiliation{Brazilian Synchrotron Light Laboratory (LNLS), Brazilian Center for Research in Energy and Materials (CNPEM), Campinas 13083-970, SP, Brazil}
	
	\author{K.~R.~Pakuszewski}
	\affiliation{Instituto de F\'{\i}sica ``Gleb Wataghin'', UNICAMP, 13083-859, Campinas, SP, Brazil}
	
	\author{C.\ Adriano}
	\affiliation{Instituto de F\'{\i}sica ``Gleb Wataghin'', UNICAMP, 13083-859, Campinas, SP, Brazil}
	
	\author{P.~G.\ Pagliuso}
	\affiliation{Instituto de F\'{\i}sica ``Gleb Wataghin'', UNICAMP, 13083-859, Campinas, SP, Brazil}
	\affiliation{Los Alamos National Laboratory, Los Alamos, New Mexico 87545, USA}

	\author{M.\ Nicklas}
	\email{Michael.Nicklas@cpfs.mpg.de}
	\affiliation{Max Planck Institute for Chemical Physics of Solids, N\"{o}thnitzer Stra{\ss}e 40, D-01187 Dresden, Germany}

\date{\today}
	
\begin{abstract}

The Weyl semimetal CeAlGe is a promising material to study nontrivial topologies in real and momentum space due to the presence of a topological magnetic phase. Our results at ambient pressure show that the electronic properties of CeAlGe are extremely sensitive to small stoichiometric variations. In particular, the topological Hall effect (THE) present in CeAlGe is absent in some samples of almost identical chemical composition. The application of external pressure favors the antiferromagnetic ground state. It also induces a THE where it was not visible at ambient pressure. Furthermore, a small pressure is sufficient to drive the single region of the THE in magnetic fields into two different ones. Our results reveal an extreme sensitivity of the electronic properties of CeAlGe to tiny changes in its chemical composition, leading to a high tunability by external stimuli. We can relate this sensitivity to a shift in the Fermi level and to domain walls.   
\end{abstract}

\maketitle

\section{INTRODUCTION}
	
The interaction between electronic charge transport and magnetization dynamics is at the basis of spintronics \cite{bader2010spintronics}. This field offers the opportunity to enable and develop new memory devices that are faster and more energy efficient than the semiconducting ones \cite{hoffmann2015opportunities}. In particular, combining spintronics with nontrivial spin textures, such as chiral domain walls or skyrmions, is a promising way to achieve improved memory devices \cite{brataas2013chiral,yang2021chiral}. The topological Hall effect (THE), which may arise from these textures, opens an improved way to manipulate the magnetization in such devices due to the interplay between Weyl fermions and domain walls through the topological Hall torque \cite{yamanouchi2022observation}.
	
An excellent playground for the study of domain walls and non-trivial topological quasiparticles, such as Weyl fermions, is the $Ln$Al$Pn$  family ($Ln$ = lanthanides, $Pn$ = Ge and Si). These materials crystallize in a noncentrosymmetric crystal structure ($I4_{1}md$), breaking the space-inversion (SI) symmetry. In addition, several members exhibit magnetic ordering at low temperatures, breaking time-reversal (TR) symmetry. The breaking of SI or TR symmetries is essential for Weyl semimetals \cite{yan2017topological,zhang2018towards,armitage2018weyl}. Several compounds in this family have been  predicted or experimentally demonstrated to be Weyl semimetals  \cite{ng2021origin,xu2017discovery,su2021multiple,sanchez2020observation,destraz2020magnetism,yang2020transition,xu2021shubnikov,gaudet2021weyl,yang2023stripe,puphal2020topological}. 

Crystalline structures without inversion symmetry can give rise to Dzyaloshinskii-Moriya interactions (DMIs), which together with spin-orbit coupling are key ingredients for chiral domain walls \cite{kimbell2022challenges}. The importance of the DMIs for the topological behavior has been established for NdAlSi \cite{gaudet2021weyl}, and  chiral domain walls have recently been observed in CeAlSi \cite{sun2021mapping}. Moreover, the scattering of Weyl fermions through domain walls in CeAlSi leads to a skew-scattering contribution in the anomalous Hall effect (AHE) \cite{piva2023topological}. This makes the members of the $Ln$Al$Pn$ family excellent candidates to study the interplay between domain walls and nontrivial topology.  
	
Here, we focus on the Weyl semimetal CeAlGe. Recently, the presence of a topological magnetic phase in CeAlGe for a magnetic field applied parallel to the $c$ axis has been reported \cite{puphal2020topological}. Previous studies have shown that ferromagnetic, antiferromagnetic, or both phases are present in samples of this material, depending on  small stoichiometric variations \cite{dhar1992magnetic,dhar1996structural,hodovanets2018single,puphal2019bulk}. The previous results indicate that the properties of this compound are tightly coupled to its crystalline structure, suggesting that small changes in the lattice parameters, for example, by the application of external pressure, can lead to large changes in its ground-state properties. An antiferromagnetic ordering close to the stoichiometric ratio of 1:1:1 was observed below $\approx 5$~K at ambient pressure \cite{puphal2019bulk}. In addition, the presence of  domain walls with high electrical resistance in CeAlGe has been proposed to be responsible for a recently observed singular angular magnetoresistance \cite{suzuki2019singular}. Moreover, the Fermi surface of CeAlGe is formed by small hole and electron pockets \cite{suzuki2019singular}. Despite the sensitivity of the magnetic structure, similar electronic band structures were found when antiferromagnetic or ferromagnetic order was considered as shown in Refs.\ \cite{he2023pressure} and \cite{chang2018magnetic}, respectively. The first study yielded 20 Weyl nodes in the antiferromagnetic phase, whereas the second reported a detailed analysis of the position of 16 nodes considering ferromagnetic order. The application of external pressure has been shown to be an effective tool for tuning Weyl points closer to the Fermi energy in closely related materials without introducing additional disorder \cite{dos2016pressure,liang2017pressure,hirayama2015weyl,rodriguez2020two} and for modifying the domain-wall landscape in its sister compound CeAlSi \cite{piva2023topological}.
 
In this paper, we present electrical transport studies under hydrostatic pressure on CeAlGe samples with slightly different compositions. Our data show that the topological Hall effect present in CeAlGe is sample dependent and extremely sensitive to small stoichiometric variations. The application of external pressure leads to the evolution of a single THE region into two distinct regions in magnetic fields applied parallel to the $c$ axis, possibly due to changes in the domain-wall landscape, in agreement with other results \cite{he2023pressure}. Remarkably, we induced a THE by applying external pressure even in samples where it was absent in the low-pressure regime. Our data highlight that CeAlGe is a highly tunable material.\\

\section{METHODS}
	
Single crystals of CeAlGe were grown by an Al-flux technique as described in Ref.\,\cite{hodovanets2018single}. The crystal structure was confirmed by powder x-ray diffraction. Two single crystals were selected for the electrical transport experiments. Their exact composition was determined by wavelength dispersive x-ray spectroscopy (WDS) performed on ten different spots on each single crystal. The WDS results yielded similar average stoichiometric ratios of Ce:Al:Ge of 0.996(1):1.034(3):0.970(2) for sample 1 (s1) and of 0.992(1):1.031(2):0.977(2) for sample 2 (s2). These ratios are the closest to 1:1:1 reported so far for Al-grown single crystals, almost comparable to floating-zone grown crystals \cite{puphal2019bulk,he2023pressure}. The only noticeable deviation from a 1:1:1 stoichiometry is found for Al and Ge, which may indicate some small site disorder between Al and Ge. Magnetization measurements were performed out using an MPMS3 (Quantum Design). Measurements of the electrical transport properties under hydrostatic pressure were conducted using a LR700 resistance bridge (Linear Research) in a physical property measurement system (Quantum Design) equipped with a 9-T superconducting magnet. Electrical transport data were collected at fixed temperatures during increasing and decreasing magnetic-field runs. The magnetic field was stabilized prior to each measurement. Pressures up to 2.4~GPa were generated utilizing a self-clamped piston-cylinder-type pressure cell. Silicon oil served as the pressure transmitting medium, and lead served as the pressure gauge.
\\

\begin{figure}[!t]
    \includegraphics[width=0.85\linewidth]{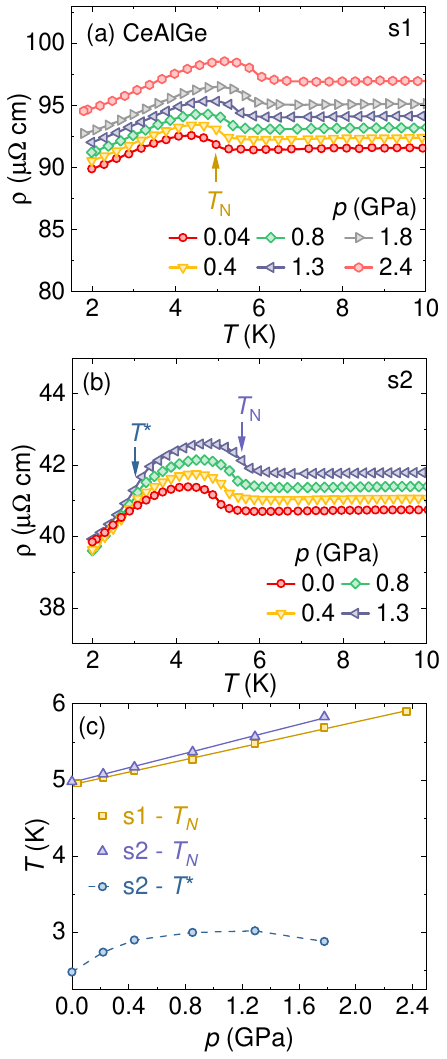}
	\caption{Electrical resistivity $\rho$ as a function of temperature at different pressures for (a) s1 and (b) s2. The difference in the absolute values in $\rho$ is due to difficulties in determining the geometric factor for s1, which did not have an ideal bar-shape geometry. (c) Temperature-pressure phase diagram. The solid lines are linear fits, and the dashed line is a guide to the eye.}
	\label{resistivity}
\end{figure}

\begin{figure}[!t]
    \includegraphics[width=0.85\linewidth]{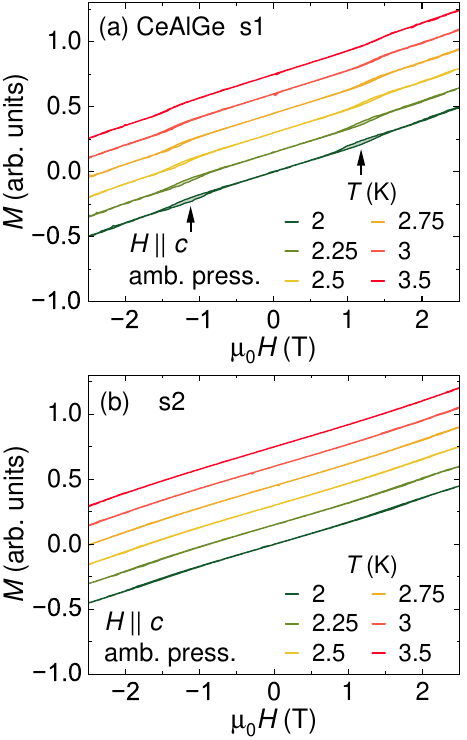}
	\caption{Magnetization $M$ as a function of magnetic-field $H\parallel c$ for (a) s1 and (b) s2 at ambient pressure for magnetic-field $-2.5{\rm~T}\leq \mu_0H \leq 2.5{\rm~T}$. The field sweeps for increasing and decreasing field were taken between $-7$ and 7~T, see Fig.~\ref{Magapp} in the Appendix. The curves have been shifted  vertically  for clarity. Arrows mark the two hysteresis regions.}
	\label{Magnetization}
\end{figure}

\section{RESULTS AND DISCUSSION}

Figures~\ref{resistivity}(a) and \ref{resistivity}(b) present the temperature dependence of the electrical resistivity $\rho$ for the two investigated CeAlGe samples at different applied pressures.  At low temperatures, an upturn in the resistivity close to 5~K, visible for both crystals, characterizes the antiferromagnetic transition temperature $T_N$, in agreement with previous reports \cite{puphal2019bulk,puphal2020topological,hodovanets2018single}. A residual resistivity ratio (RRR) of 2.46 for s1 and  of 2.59 for s2 indicates a comparable quality of both samples. However, s2 displays an additional shoulder below $T_N$, labeled with $T^*$ in Fig.~\ref{resistivity}(b), which could be related to the presence of another magnetic phase as previously reported for CeAlGe \cite{puphal2019bulk}. The application of external pressure favors the antiferromagnetic state in both samples as clearly seen in the temperature-pressure phase diagram shown in Fig.~\ref{resistivity}(c). The antiferromagnetic transition temperature $T_{N}(p)$ increases linearly with a rate of  $dT_{N}(p)/dT\approx0.41(1)$~K/GPa for s1 and $0.48(1)$~K/GPa for s2 with the latter being only slightly larger. The second anomaly, observed only in s2, shows a domelike shape as a function of pressure. Its origin is not yet clear, but it is likely to be of magnetic origin.

Magnetization ($M$) data as a function of the applied magnetic-field $H\parallel c$ at ambient pressure are presented in Figs.~\ref{Magnetization}(a) and \ref{Magnetization}(b) for s1 and s2, respectively. Sample 1 displays a small hysteresis in $M(H)$ in a narrow region around 1.2~T, which is absent in s2. No hysteresis is found above 3~K. We note that the hysteretic behavior is observed in the same region where neutron-diffraction experiments detected a transition from an incommensurate to a ferromagnetic polarized phase in CeAlGe \cite{puphal2020topological}.

\begin{figure}[!t]
	\includegraphics[width=0.85\linewidth]{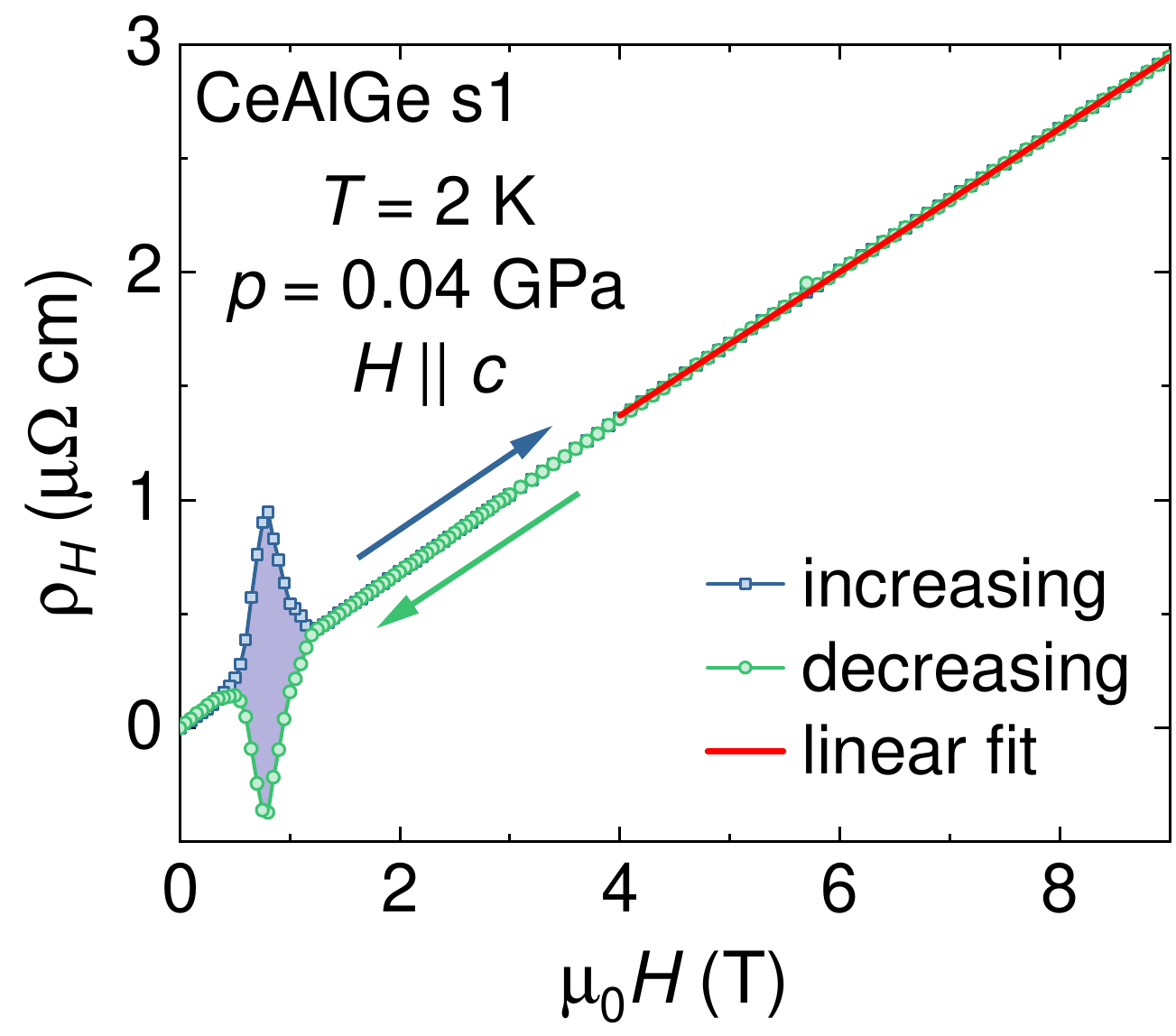}
	\caption{Example data of the Hall resistivity $\rho_{H}(H)$ for increasing and decreasing magnetic fields to illustrate the procedure for extracting the topological Hall effect $\Delta \rho_{H} /2$. The red line is a linear fit to determine the ordinary Hall constant $R_H$.}
	\label{deltarho_s1_explanation}
\end{figure}

We now turn our attention to the evolution of the Hall resistivity $\rho_{H}$ of CeAlGe as a function of the magnetic field applied parallel to the $c$ axis. As can be seen from the ambient pressure data in Fig.~\ref{deltarho_s1_explanation}, at 2~K a maximum (minimum) appears at $\approx$ 0.8~T in $\rho_{H}$, after being antisymmetrized with respect to the origin for measurements performed with increasing (decreasing) magnetic fields. These features occur in the same field range as the topological magnetic phase that has been observed in CeAlGe before \cite{puphal2020topological}. For magnetic fields higher than 2~T, $\rho_{H}(H)$ displays a linear behavior.

The total Hall resistivity can be written as $\rho_{H} = \rho^{\rm OHE} + \rho^{\rm AHE} + \rho^{\rm THE}$, where $\rho^{\rm OHE}$ is the ordinary Hall effect caused by the Lorentz force acting on the charge carriers described by $\rho^{\rm OHE} = R_{H} \mu_{0} H$, where $R_{H}$ is the Hall coefficient, $\mu_{0}$ is the magnetic constant, and $H$ is the applied magnetic field. The second term $\rho^{\rm AHE} = R_{s} \rho^2 M$ is the anomalous Hall effect. It arises due to the presence of magnetism, strong spin-orbit coupling, and/or from an intrinsic Berry curvature in the momentum space \cite{kimbell2022challenges}, and it is proportional to the bulk magnetization $M$ and $\rho^2$. $R_{s}$ is the anomalous Hall coefficient. The last term $\rho^{\rm THE}$ is the topological Hall effect caused by the presence of a Berry phase curvature created by topological spin textures, such as chiral domain walls or skyrmions \cite{kimbell2022challenges}. It is worth noting that both ordinary and anomalous Hall effects do not depend on the history of the measurement. The tiny hysteresis in $M(H)$ that is observed in s1 [see Fig.~\ref{Magnetization}(a)] does not give rise to any visible feature in $\rho_{H}$, in agreement with previous reports \cite{suzuki2019singular,hodovanets2018single,he2023pressure}. We note that a large AHE was observed in a previous study \cite{puphal2020topological}. Since the AHE is proportional to the longitudinal resistivity, high quality samples with lower residual resistivity could have a weaker contribution of the AHE to the Hall effect. In fact, the AHE was observed in CeAlGe for samples with an RRR of 1.3 \cite{puphal2020topological}, and it is absent for samples with an RRR higher than 1.85 \cite{hodovanets2018single,he2023pressure}.

On the other hand, the THE can be caused by chiral domain walls or other unconventional spin textures, which may depend on the history of the applied magnetic fields. Therefore, a simple way to extract the THE is to subtract the antisymmetrized Hall resistivity data obtained in experiments with decreasing applied magnetic field (9 $\rightarrow$ $-9$~T) from those recorded with increasing the applied magnetic field ($-9$ $\rightarrow$ $9$~T), and then divide the result by 2 to scale the absolute values, which we will denote hereafter as $\Delta \rho_{H}/2$.

\begin{figure}[!t]
	\includegraphics[width=0.85\linewidth]{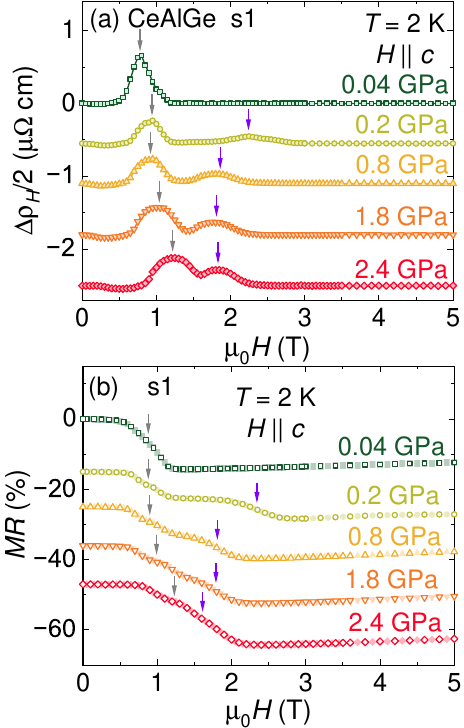}
	\caption{(a) $\Delta \rho_{H} /2$ and (b) ${\rm MR}$ as functions of magnetic-field $H\parallel c$ at different pressures at 2~K for s1. The curves have been shifted vertically for better visualization. The open (transparent) symbols in (b) are obtained with increasing (decreasing) magnetic field. The arrows in panels (a) and (b) mark the maxima in $\Delta \rho_{H} /2$ and the midpoints of the step-like feature in ${\rm MR}(H)$, respectively.}
	\label{deltarho_s1}
\end{figure}

Figure~\ref{deltarho_s1}(a) presents $\Delta \rho_{H}(H)/2$ at different pressures at 2~K for sample 1. At ambient pressure, a peak at 0.8~T is clearly visible. A small increase in the externally applied pressure of only 0.2~GPa is sufficient to split the peak into two, the first occurring around 0.9~T, and the second one occurring around 2.2~T. At the same time, the THE signal is suppressed. Further increase in pressure enhances the height of the second peak and drives both peaks closer together, reaching 1.2 and 1.8~T for the position of the first and second maxima in $\Delta \rho_{H}/2$ at 2.4~GPa, respectively. We note that this response is in contrast to a previous report \cite{he2023pressure} in which the peak in the Hall resistivity splits only at 1.8~GPa, and where further increase in pressure drives the peaks further apart. This may be taken as another indication of the strong sensitivity of the properties of CeAlGe to small differences in the sample stoichiometry. 

To gain further insight into the evolution of the magnetic properties with pressure, we now turn to the magnetoresistance ${\rm MR}=[\rho(H)-\rho(H=0)]/\rho(H=0)$  at 2~K with the field applied parallel to the $c$ axis [see Fig.~\ref{deltarho_s1}(b)]. At ambient pressure, a steplike feature is observed at 0.9~T. Furthermore, there is no difference in the measurements performed with increasing and decreasing fields, in contrast to what is observed for the Hall resistivity. The application of external pressure splits the steplike anomaly in the ${\rm MR}(H)$ into two separate steps, similar to the splitting of the maximum in $\Delta \rho_{H}/2$. The steplike feature at 0.9~T at ambient pressure splits already at 0.2~GPa into two steps, the first one occurring at 0.9~T and the second one occurring at 2.4~T. Again, a further increase in pressure tends to merge the steps at 1.2 and 1.6~T into a single step at 2.4~GPa. Note that the total step size in the MR, about $12\%$ between 0 and 5~T, remains almost the same between ambient pressure and 0.2~GPa. This is a hint that the single steplike feature at ambient pressure actually consists of two overlapping steps that are driven apart by the application of pressure. The positions of the features in $\Delta \rho_{H}(H)/2$ and ${\rm MR}(H)$ show a similar evolution as a function of pressure [see the arrows in Figs.~\ref{deltarho_s1}(a) and \ref{deltarho_s1}(b)]. This suggests that the THE in CeAlGe is indeed related to magnetic phases or structures that merge at ambient pressure and can be separated by slight compression of the material.

\begin{figure*}[!t]
	\includegraphics[width=0.80\linewidth]{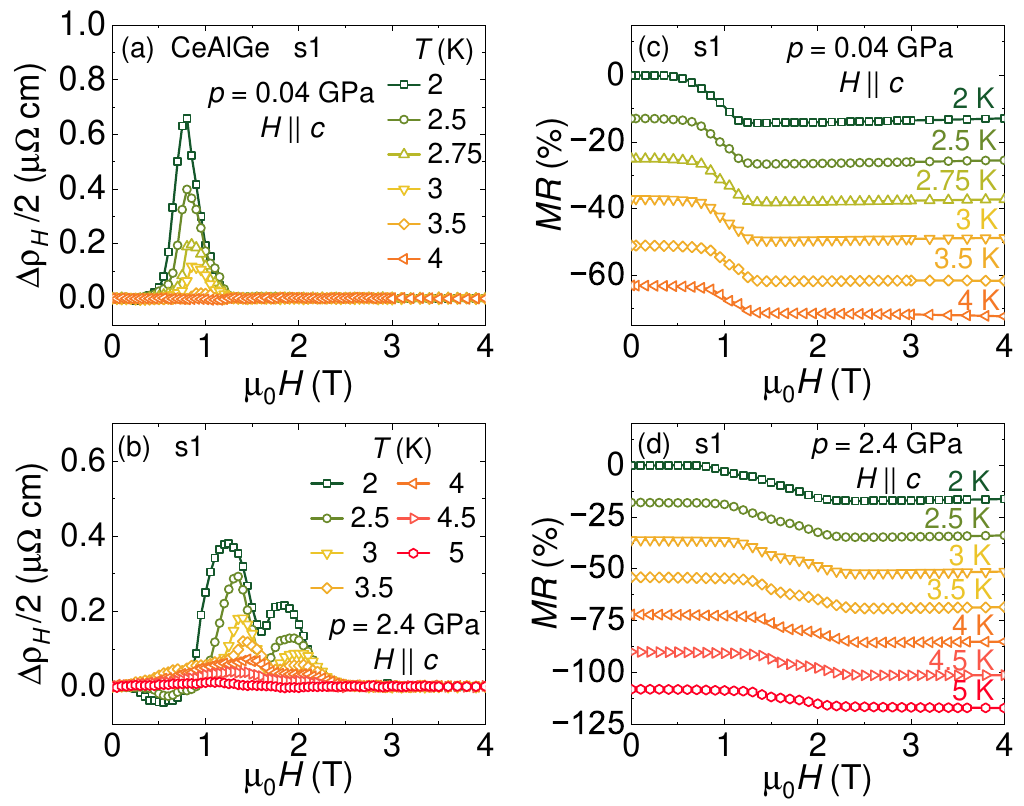}
	\caption{(a) and (b) Temperature evolution of the field dependence of the $\Delta \rho_{H}(H) /2$ and (c) and (d) ${\rm MR}$ taken on s1 at ambient pressure and 2.4~GPa. The magnetic field was applied parallel to the $c$ axis in all panels. The curves shown in panels (c) and (d) have been shifted vertically for better visualization.}
	\label{THExT_s1}
\end{figure*}

The connection of the THE with the presence of the magnetic phase(s) is also shown by the temperature evolution of the peak in $\Delta \rho_{H}(H)/2$ [see Figs.~\ref{THExT_s1}(a) and \ref{THExT_s1}(b)]. The size of the maximum in $\Delta \rho_{H}(H)/2$ decreases with increasing temperature and disappears close to the N\'eel temperature $T_{N}$. On the other hand, the steplike features in the magnetoresistance, Figs.~\ref{THExT_s1}(c) and \ref{THExT_s1}(d), are much more robust to increasing temperature and are clearly visible at temperatures higher than the peak in $\Delta \rho_{H}(H)/2$, but also disappear close to $T_N$. We may speculate that the origin of the different behavior lies in the different response of the MR and the Hall effect to the evolution of the magnetic domain structure with temperature. It has been shown that a THE can be generated by chiral domain walls or by the scattering of Weyl fermions through the domain walls \cite{sorn2021domain,piva2023topological}. It is, therefore, conceivable that the features in $\Delta \rho_{H}(H)/2$ that may arise from the domain walls are less robust to increasing temperature than the MR features associated with the bulk domains. For example, a complex evolution of the domains and domain walls with increasing temperature from 2~K up to the magnetic transition temperature has been observed in the sister compound CeAlSi in magneto-optical Kerr effect experiments \cite{sun2021mapping}.

\begin{figure}[!t]
	\includegraphics[width=0.85\linewidth]{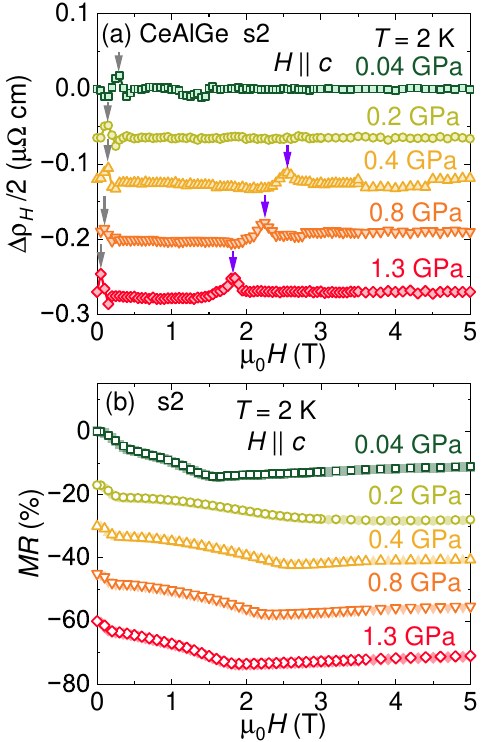}
	\caption{(a) and (b) Pressure evolution of the field dependence of $\Delta \rho_{H}(H) /2$ and ${\rm MR}$ at 2~K for s2. The open (transparent) symbols in (b) were obtained with increasing (decreasing) magnetic field. The magnetic field was applied parallel to the $c$ axis in all plots. The curves have been shifted vertically for better visualization.}
	\label{THExT_s2}
\end{figure}

To learn more about the sensitivity of the THE in CeAlGe to small changes in the sample stoichiometry, we come to s2. Sample 2 has almost the same average stoichiometric ratio and only a slightly larger RRR than s1, suggesting the presence of similar magnetic and electrical transport properties. Contrary to the expectation, we find distinct differences in s2 compared to s1. We analyzed the Hall resistivity in the same way as we did for s1, but at ambient pressure, we find no peak in $\Delta \rho_{H}(H)/2$ in a similar field as in s1 at 2~K [Fig.~\ref{THExT_s2}(a)]. However, a small peak is visible at 0.3~T, a much smaller field than the the feature in s1. It shifts monotonically to lower fields with increasing pressure. At 0.4~GPa, a peak appears at 2.6~T. The position of this peak is very close to that of the second peak in s1 at 0.2~GPa. Similar to the behavior in s1, the application of pressure pushes the peak toward lower fields, suggesting a similar origin. We emphasize that we do not find a similar peak at the position of the first peak seen in s1, even under pressure.

\begin{figure}[!t]
	\includegraphics[width=0.85\linewidth]{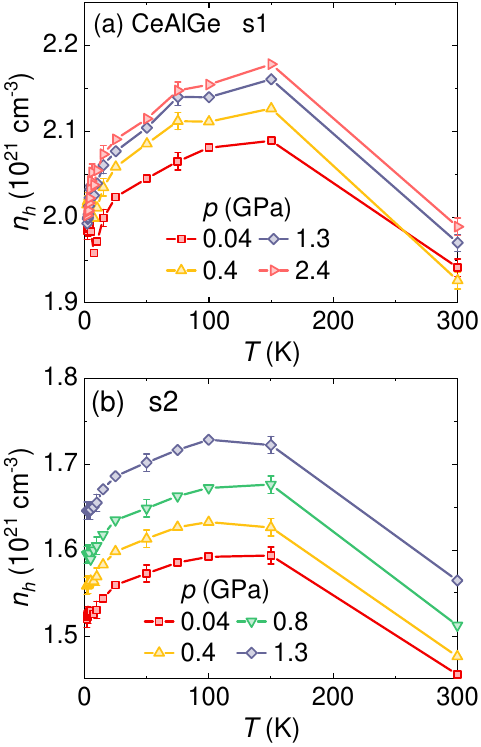}
	\caption{Hole density ($n_{h}$) as a function of temperature at different pressures for (a) s1 and (b) s2, respectively.}
	\label{carriers}
\end{figure}

The magnetoresistance provides additional information. Although the ${\rm MR}$ for s1 shows a clear steplike structure and well-defined plateaus, these features are washed out for s2 [Fig.~\ref{THExT_s2}(b)]. This makes it more difficult to associate features in the MR with corresponding signatures in the topological Hall effect. At low field, a prominent feature in ${\rm MR}(H)$ tracks the peak in $\Delta \rho_{H}(H) /2$. The broad washed-out step at higher fields is already present at ambient pressure where no peak in $\Delta \rho_{H}(H) /2$ can be resolved; at higher pressures, the onset of the step in ${\rm MR}(H)$ appears to track the position of the peak in $\Delta \rho_{H}(H) /2$. This result for sample 2 compared to the clear data for s1 is surprising, since our WDS analysis indicates almost the same sample stoichiometry, and the higher RRR even suggests that s2 should be of slightly \textit{better} quality than s1.

The origin of the THE can be related to the presence of a topologically nontrivial magnetic texture in CeAlGe \cite{puphal2020topological}. In the sister compound CeAlSi, the presence of an anomalous Hall effect and a loop Hall effect could be explained by the scattering of the Weyl fermions through domain walls \cite{piva2023topological}. Chiral domain walls have also been observed experimentally  \cite{sun2021mapping}, and their sensitivity to magnetoelastic couplings has been established in CeAlSi \cite{xu2021picoscale}. In CeAlGe, on the other hand, high-resistance domain walls have been shown to give rise to a singular angular magnetoresistance as a function of the in-plane magnetic field \cite{suzuki2019singular}. Therefore, it is reasonable to speculate that the domain walls also play an important role in the transport properties of CeAlGe for the field along the $c$ axis. A similar conclusion was reached in Ref.~\cite{he2023pressure}. However, we find different results at the first sight for almost identical samples. In the following, we attempt to reconcile our data on the THE with the results reported in the literature \cite{hodovanets2018single,puphal2020topological,he2023pressure} and to develop a common picture.

By analyzing the ordinary Hall effect, we can get information about the charge-carrier density and possible differences between the two samples. The Hall response at high fields is linear with field, indicating that it is dominated by the ordinary Hall effect (see Fig.~\ref{deltarho_s1_explanation}). From a linear fit to $\rho_H(H)$ in the range of 4 to 9~T, as highlighted by the red solid line in Fig.~\ref{deltarho_s1_explanation}, we can determine the ordinary Hall constant $R_H$ and, subsequently, the carrier-density $n_{h}$. We find a hole like behavior in the whole temperature and pressure range. Figures~\ref{carriers}(a) and \ref{carriers}(b) present $n_{h}$ as a function of temperature at different pressures for s1 and s2, respectively. At ambient pressure  we find $n_{h}=1.94(1) \times 10^{21}$ and $n_{h}=1.46(1) \times 10^{21}$~holes/cm$^{3}$ for s1 and s2 at room temperature, respectively. In both samples, $n_{h}$ shows a smooth variation as a function of temperature, only at low temperatures a pronounced kink appears related to the magnetic order. At 2~K, $n_{h}$  reaches $1.98(1) \times 10^{21}$ and $1.52(1) \times 10^{21}$~holes/cm$^{3}$ for s1 and s2, respectively. The application of external pressure leads to a monotonic increase in the hole like carrier density for both samples within the error bars of the analysis. We note that we find a considerably higher carrier density for the two investigated samples than reported in Refs.~\cite{suzuki2019singular,hodovanets2018single,he2023pressure}. We will return back to this point later.

The difference in the carrier density between the two samples is remarkable given that both samples are from the same growth and have almost identical stoichiometry and a similar RRR. At ambient pressure, $n_{h}$ of S1 is around 25~\% larger than that of S2. This may indicate a different position of the Fermi level in the band structure of the two samples, possibly caused by the different levels of defects present in s1 and s2, implying a different distance of the Weyl nodes from the Fermi level as also observed in the sister compound CeAlSi \cite{yang2021noncollinear}. This is consistent with a scenario where the Weyl nodes are closer to the Fermi surface in s1 than in s2, leading to the observation of more pronounced THE features in s1 than the pressure-induced THE in s2.

Previous reports show apparently conflicting data on the Hall resistivity of CeAlGe where a THE has been observed \cite{puphal2020topological,he2023pressure} or where it has not been reported \cite{hodovanets2018single,suzuki2019singular}.
Compared to our samples, all available data in the literature find a significantly lower carrier density. Neglecting the influence of disorder scattering, this is also consistent with our samples having the highest RRR. Our material is almost as close to the 1:1:1 stoichiometry as the floating zone grown samples \cite{puphal2020topological}. These samples have the lowest reported carrier density and are the only ones to show a pronounced AHE. However, they have a THE in a similar pressure range as our sample 1. Although a THE cannot be identified in the data of Suzuki {\it et al.}\ \cite{suzuki2019singular}, it was overlooked in Hodovanets {\it et al.}\ \cite{hodovanets2018single}. In the latter work, a shallow minimum can be identified at about 2~T. This seems to suggest that the THE is a universal feature in CeAlGe, noting the large deviations from the 1:1:1 composition. 

Coming back to our data taken on two high quality samples with almost identical chemical compositions close to 1:1:1 and RRR around 2.5, which show a qualitatively different Hall response. This demonstrates the extreme sensitivity of the electronic properties of CeAlGe. Although s2 does not show a THE at ambient pressure, we can induce it by applying pressure. Our results together with those reported in the literature indicate the robustness of THE in CeAlGe. We note that the THE phase appears in different magnetic-field ranges in our data and in the data sets reported in the literature. Furthermore, the THE regime shows a different pressure evolution in our data and in those of Ref.~\cite{he2023pressure}. It is important to emphasize that all data suggest a much broader field range of the topologically nontrivial magnetic structure than the region where the THE is observed. This points to the importance of the domain walls and not (only) the topologically nontrivial magnetic structure as the origin of the THE. A field-driven modification of the domain-wall structure is easily conceivable and likely to be more sensitive to lattice defects. Unfortunately, it is not straightforward to quantitatively relate the observed behavior to the sample composition and its structural properties. This requires further studies and seems to be very challenging at the resolution level required.\\

\section{CONCLUSION}
\nobreak
In conclusion, our study shows that the low-temperature properties of CeAlGe are extremely sensitive to small stoichiometric variations especially the topological Hall effect. Simultaneous measurements of the magnetoresistance and the Hall effect suggest that inhomogeneities in the topological magnetic phase, i.e., a different domain structure, have a strong influence on the observed THE. Despite the only small stoichiometric variations between the samples, we also find a large difference in the carrier density, suggesting a significantly different position of the Fermi level in the band structure in different samples. This can also directly change the contribution of Weyl fermions to the observed transport properties either by shifting the Weyl cones with respect to the Fermi energy and/or by a smaller or larger contribution of normal electrons to the transport properties. The latter should not affect the size of the THE. Finally, our study shows that a consistent data set on different samples using different physical probes, would be highly desirable to gain a complete understanding of the topological properties of CeAlGe.\\

%\end{document}

%\section*{DATA AVAILABILITY}
Data that underpin the ﬁndings of this paper are available at Edmond – the open research data repository of the Max Planck Society at Ref.~\cite{EDMOND}.

\begin{acknowledgments}
We acknowledge fruitful discussions with L. O. Kutelak. We thank U.\ Burkhardt for carrying out energy and wavelength dispersive x-ray analysis of the samples. This project has received funding from the European Union’s Horizon 2020 Research and Innovation Programme under the Marie Sk\l{}odowska-Curie Grant Agreement No. 101019024. This work has also been supported by the S\~ao Paulo Research Foundation (FAPESP) Grants No. 2017/10581-1, No. 2018/11364-7, No. 2020/12283-0, CNPq Grants No. 304496/2017-0, No. 310373/2019-0, and CAPES, Brazil. Work at Los Alamos was supported by the Los Alamos Laboratory Directed Research and Development program through Project No. 20210064DR.
\end{acknowledgments}

\section*{APPENDIX: Magnetization measurements} 

%\section*{Magnetization measurements}

\begin{figure}[!t]
    \includegraphics[width=0.85\linewidth]{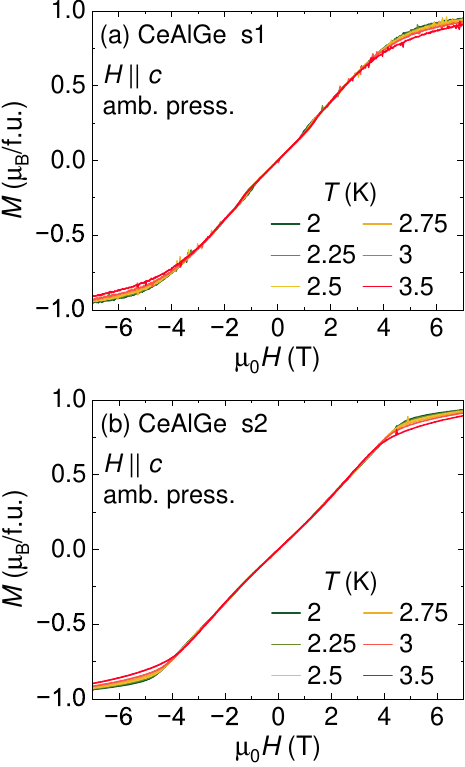}
	\caption{Magnetization ($M$) as a function of the magnetic field $H\parallel c$ for (a) s1 and (b) s2 at ambient pressure.}
	\label{Magapp}
\end{figure}

Figure~\ref{Magapp} presents magnetization data as a function of the magnetic field at ambient pressure for CeAlGe s1 and s2 for fields in the full investigated field range $-7{\rm~T}\leq \mu_0H \leq 7{\rm~T}$. 

\bibliography{CeAlGe}

\end{document}